\documentclass{article}

\usepackage{PRIMEarxiv}

\usepackage[utf8]{inputenc} 
\usepackage[T1]{fontenc}    
\usepackage{hyperref}       
\usepackage{url}            
\usepackage{booktabs}       
\usepackage{amsfonts}       
\usepackage{nicefrac}       
\usepackage{microtype}      
\usepackage{lipsum}
\usepackage{fancyhdr}       
\usepackage{graphicx}       
\graphicspath{{media/}}     
\usepackage{amsmath}

\title{Multi-Task Multi-Agent Shared Layers are Universal Cognition of Multi-Agent Coordination
}

\author{
  Jiawei Wang \\
  Tongji University \\
  \texttt{wangjw@tongji.edu.cn} \\
   \And
  Jian Zhao \\
  Polixir \\
  \texttt{jian.zhao@polixir.ai} \\
  \And
  Zhengtao Cao \\
  Polixir \\
  \texttt{zhengtao.cao@polixir.ai} \\
  \And
  Ruili Feng \\
  Alibaba Group \\
  \texttt{ruilifengustc@gmail.com} \\
  \And
  Rongjun Qin \\
  Nanjing University, Polixir \\
  \texttt{qinrj@lamda.nju.edu.cn} \\
  \And
  Yang Yu* \\
  Nanjing University, Polixir \\
  \texttt{yuy@nju.edu.cn} \\
}

\begin{document}
\maketitle

\begin{abstract}
Multi-agent reinforcement learning shines as the pinnacle of multi-agent systems, conquering intricate real-world challenges, fostering collaboration and coordination among agents, and unleashing the potential for intelligent decision-making across domains.
However, training a multi-agent reinforcement learning network is a formidable endeavor, demanding substantial computational resources to interact with diverse environmental variables, extract state representations, and acquire decision-making knowledge.
The recent breakthroughs in large-scale pre-trained models ignite our curiosity: Can we uncover shared knowledge in multi-agent reinforcement learning and leverage pre-trained models to expedite training for future tasks?
Addressing this issue, we present an innovative multi-task learning approach that aims to extract and harness common decision-making knowledge, like cooperation and competition, across different tasks. 
Our approach involves concurrent training of multiple multi-agent tasks, with each task employing independent front-end perception layers while sharing back-end decision-making layers. 
This effective decoupling of state representation extraction from decision-making allows for more efficient training and better transferability.
To evaluate the efficacy of our proposed approach, we conduct comprehensive experiments in two distinct environments: the StarCraft Multi-agent Challenge (SMAC) and the Google Research Football (GRF) environments.
The experimental results unequivocally demonstrate the smooth transferability of the shared decision-making network to other tasks, thereby significantly reducing training costs and improving final performance. 
Furthermore, visualizations authenticate the presence of general multi-agent decision-making knowledge within the shared network layers, further validating the effectiveness of our approach.
\end{abstract}


\section{Introduction}
\label{sec:intro}
Cooperative multi-agent reinforcement learning (MARL) has garnered significant attention due to its vast potential in solving real-world challenges, including traffic light control~\cite{tralc}, autonomous cars~\cite{cars}, and robot swarm control~\cite{robots}.
However, training a multi-agent network can be prohibitively expensive. It requires frequent communication with complex environmental variables and consumes substantial time and resources.
For instance, AlphaStar \cite{vinyals2019grandmaster}, OpenAI Five \cite{berner2019dota}, and JueWu \cite{ye2020towards} required hundreds of thousands of cores and several months to achieve satisfactory performance.
Such computational demands pose barriers to general research, limiting the full utilization of this exciting progress in modern machine intelligence by academic institutions and the broader intelligence community. 
Consequently, exploring methods to effectively extract and leverage shared knowledge in multi-agent decision tasks has emerged as a prominent research direction.

Inspired by the remarkable advancements of large-scale pre-trained models, a fascinating and promising question arises: Can we separate shared knowledge from the multi-agent learning framework and store it in pre-trained networks to reduce the cost of task transfer?

The primary challenge in applying pre-trained models to multi-agent reinforcement learning arises from the frequent variations in the formats of state and action spaces, even within a single task.
Pre-trained models typically require standardized input and output formats, such as image pixels in computer vision or text tokens in natural language processing.
Fortunately, this paper successfully identifies unique shared knowledge in multi-agent reinforcement learning. 
Agents often exhibit shared decision-making priors for specific tasks, such as cooperation and competition, while certain global conditions influence the decisions of all agents across diverse scenarios and tasks.
However, extracting and sharing this common decision-making knowledge is not a straightforward path.
In decision tasks, neural networks serve dual roles in perception and decision-making, and these functions are tightly intertwined.
Separating them and distilling a rich and universal decision-making knowledge across multi-agent tasks is a challenging endeavor.

This paper introduces a pioneering multi-task learning approach that effectively tackles the aforementioned challenges and extends the power of pre-training to the realm of multi-agent reinforcement learning.
By employing task-specific front-end perception modules and a shared back-end decision-making module, this approach effectively decouples the decision process, enabling the efficient disentanglement and extraction of shared decision-making knowledge.
To handle varying action spaces, actions are positioned at the input end of the network, ensuring consistency in network output dimensions, which is referred to as the action prepositioning network (APN).
Additionally, dynamic adaptive weights for multi-task losses are introduced to enhance multi-task pre-training efficiency.

We conduct extensive experiments in StarCraft Multi-agent Challenge (SMAC)~\cite{samvelyan2019starcraft} and Google Research Football (GRF)~\cite{kurach2020google} environments.
Our findings demonstrate the effectiveness of our approach in extracting common decision knowledge across different tasks and efficiently transferring it to new tasks, significantly reducing training costs and improving network performance.
Our method significantly benefits from increasing the number of pre-training tasks, showing potential in large-scale scenarios.
Furthermore, through visualization, we validate that the extracted decision networks contain universal decision knowledge for multi-agent coordination, shedding new light on future research of pre-trained models in multi-agent reinforcement learning.

\section{Related Work}
\label{sec:related_work}

\subsection{Pre-trained Model}

A pre-trained model is a model that's already well-trained, and it shares its knowledge with a new model to speed up the new model's training, without the need to start from scratch. 
Pretrained models~\cite{weiss2016survey,zhuang2020comprehensive} extract common knowledge of a broad range of tasks from data. It can accelerate and improve training of downstream tasks, by leveraging the common knowledge stored in them in various means.
Pre-trained models can be applied to a wide range of downstream tasks, such as image classification~\cite{brock2018large, He_2020_CVPR, chen2020simple, targ2016resnet}, object detection~\cite{bochkovskiy2020yolov4,carion2020end,ren2015faster,tan2020efficientdet,zhang2021fairmot}, text classification~\cite{devlin2018bert,zhang2019ernie,sanh2019distilbert,jiao2019tinybert,lan2019albert} and more. When training for downstream tasks, we need to load the parameters of the pre-trained model's feature layers. After that, keeping the feature layers (the initial layers) fixed and fine-tuning the parameters of the later layers~\cite{yosinski2014transferable} because the initial layers primarily focus on extracting generic information. For instance, in ResNet, the early layers are responsible for capturing low-level features like textures and shapes.

\subsection{Pre-trained Model by SL}
Traditionally, pre-trained models are usually trained through supervised learning (SL) on large data sets.
They possess extensive language and visual knowledge, making them directly applicable to new tasks or amenable to fine-tuning for specific tasks.
Using pre-trained models significantly reduces the time and computational resources required to build and train models from scratch, especially when data for new tasks is limited. In supervised learning, especially in the field of image processing, we often use large labeled datasets like ImageNet~\cite{deng2009imagenet}, Open Images dataset~\cite{krasin2017openimages}, and COCO~\cite{lin2014microsoft} to pre-train models and obtain large pre-trained models such as AlexNet~\cite{krizhevsky2012imagenet}, VGG~\cite{simonyan2014very}, and ResNet~\cite{he2016deep}.
These pre-trained models typically learn generic features that can be useful for many downstream tasks.
These features speed up the process of learning specific characteristics for the downstream tasks, making the overall training faster. 

\subsection{Pre-trained Model in SSL}

Because the cost of data annotation is too high, researchers use self-supervised learning (SSL) methods to train pre-trained models.
Up to now, self-supervised pre-trained models are being applied in various domains, including NLP~\cite{devlin2018bert,radford2018improving,radford2019language, brown2020language}, computer vision~\cite{park2020contrastive,caron2020unsupervised,he2020momentum,tian2021understanding}, and audio~\cite{spijkervet2021contrastive}.
Unlike supervised learning, self-supervised pre-training doesn't require labeled data.
There are two main approaches: one involves creating self-generated tasks to learn feature representations from the data, such as tasks like image completion, text generation, image rotation, and image augmentation.
The other approach uses contrastive learning to bring similar samples closer in feature space, thus capturing common underlying features among similar samples.

\subsection{Pre-trained Model in RL}

Inspired by large-scale language modeling, DeepMind has applied a similar approach to create a unified intelligent agent called Gato~\cite{reed2022generalist}.
Gato has been trained on over $600$ different tasks, each with varying patterns, observations, and action guidelines. However, Gato, despite integrating a variety of decision tasks, still lacks generalization abilities. This is because in the single-agent domain, different decision tasks require completely different knowledge. For example, in Atari games, tasks like pong and breakout have no shared knowledge between them.
In contrast, in multi-agent tasks, such as $3s5z$ to $3s5z vs 3s6z$ in SMAC, there exists strong shared knowledge, such as teammate attributes, enemy attributes, strategies like focus fire, hit and run, and more.
In multi-agent reinforcement learning, there typically exists significant variation in the state space among different tasks.
Current pre-trained methods yield universal perceptual models that cannot be shared across diverse multi-agent tasks.
Different multi-agent tasks often involve more shared decision knowledge, such as agent cooperation. Therefore, we propose a multi-task, multi-agent pre-training approach aimed at acquiring pre-training models that encompass shared decision knowledge.

\section{Preliminaries}\label{sec:preliminary}


We model multi-agent reinforcement learning (MARL) as a decentralized partially observable Markov decision process (Dec-POMDP)~\cite{bernstein2002complexity}. A multi-agent Dec-POMDP can be represented as a tuple $(\mathcal{N},\mathcal{S},\mathcal{A},P,R,\mathcal{O},G,\gamma)$. Here, $\mathcal{N} = \{1,2,\cdots,N\}$ denotes the $N=|\mathcal{N}|$ agents in the multi-agent system. $\mathcal{S}$ represents the state space, $\mathcal{A}$ is the action space, and $\mathcal{O}$ corresponds to the observation space. The reward function is denoted as $R: \mathcal{S} \times \mathcal{A} \mapsto \mathbb{R}$. At each time step $t$, each agent $i$ performs an action $a_{i} \in \mathcal{A}$, collectively forming a joint action $\mathbf{a}$. After taking action $\mathbf{a}$ in state $s$, all agents receive rewards $R(s,\mathbf{a})$. $P$ represents the state transition probability that satisfies the Markov property: $\mathcal{S} \times \mathcal{A} \times \mathcal{S} \mapsto [0,1]$. Each agent $i$, after taking action $a_{i}$ in state $s$, observes $o_{i}$ based on the observation function $G(s, \mathbf{a}): \mathcal{S} \times \mathcal{A} \mapsto \mathcal{O}$. $\gamma \in [0, 1)$ is the discount factor. Each agent $i$ has a policy $\pi^{i}(a_{t}^{i}|o_{1:t}^{i})$ to predict action $a_{t}^{i}$ based on its historical observations. $\pi(s_{t},\mathbf{a}_{t})$ denotes the joint policy for all agents. The ultimate goal is to learn an optimal joint policy that maximizes the discounted expected return $\mathbb{E}[\sum_{t=0}^{\infty}\gamma^{t}R(s_{t},\mathbf{a}_{t})]$.

Value-based MARL algorithms, such as VDN~\cite{sunehag2017value} and QMIX~\cite{rashid2020monotonic}, collaborate by utilizing action-value functions and joint action-value functions to make optimal decisions that maximize the cumulative rewards for the entire group of agents. VDN endeavors to decompose the total action-value function, denoted as $Q_{tot}(\mathbf{s},\mathbf{a})$, into a composition of distinct local action-value functions $Q_{i}(s_{i},a_{i})$:
\begin{equation}
    Q_{tot}(\mathbf{s},\mathbf{a}) = \sum_{i=1}^{N}Q_{i}(s_{i},a_{i}),
\end{equation}

\noindent where $Q_{i}(s_{i},a_{i})$ characterizes the action-value function of agent $i$.
In the training phase, the optimization of the policy is guided by $Q_{tot}$.
Simultaneously, each agent $i$ derives its local action-value function $Q_{i}$ from the global $Q_{tot}$ value to make decisions.
QMIX operates under the assumption that a global $arg max$ performed on $Q_{tot}$ is equivalent to a set of individual $arg max$ operations performed on each $Q_{i}$.
To facilitate this, QMIX employs a mixing network as a functional expression to amalgamate $Q_{i}$ values into the $Q_{tot}$ value, enabling it to meet:
\begin{equation}
    \frac{\partial Q_{tot}}{\partial Q_{i}} \ge 0, \forall i \in \{1,2,\cdots,N\}.
\end{equation}

\section{Method}
\label{sec:method}

\begin{figure*}[ht!]
    \centering
    \includegraphics[width=0.98\linewidth]{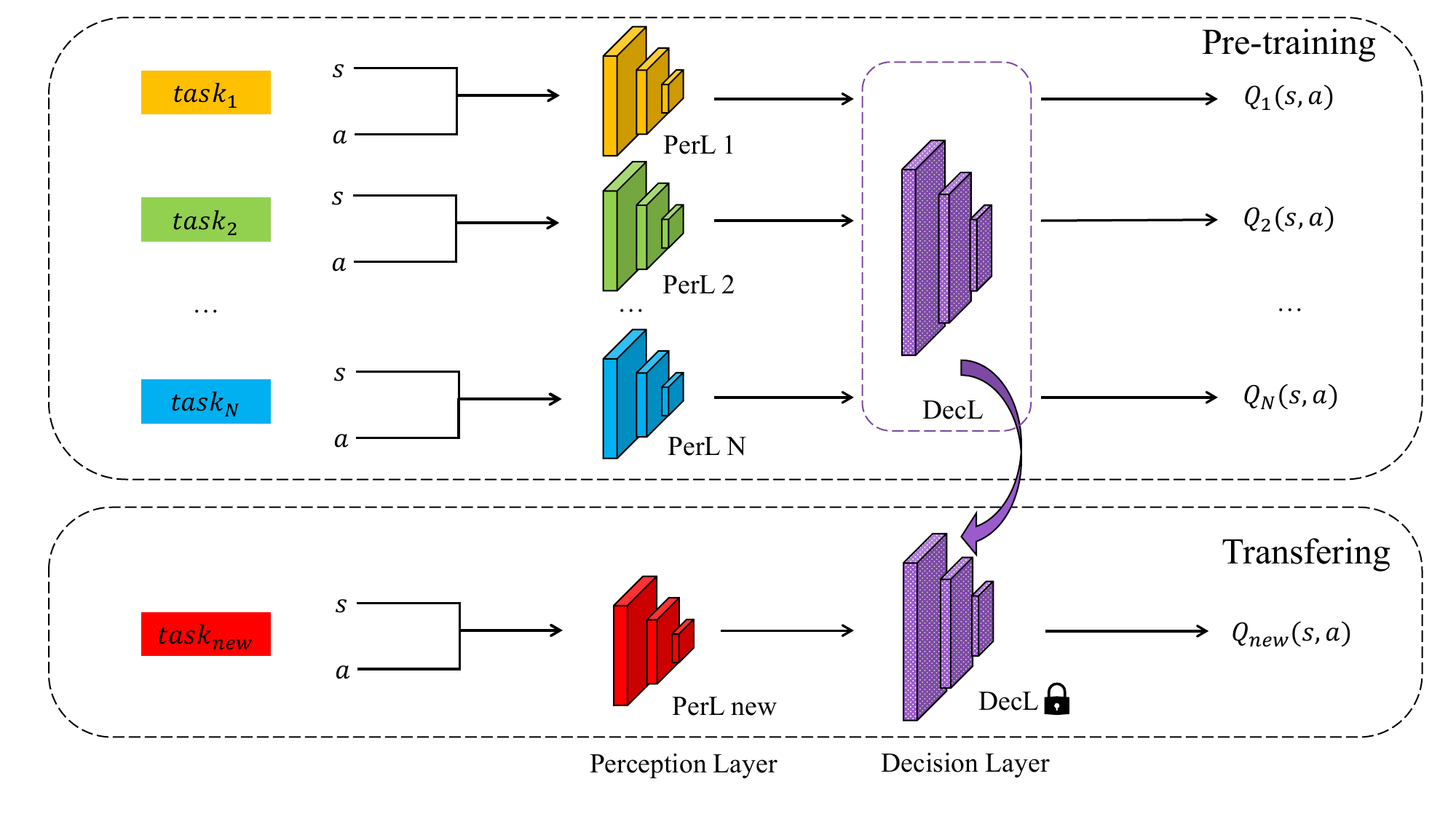}
    \caption{Overall framework of the proposed multi-task multi-agent pre-training method. The ‘lock’ icon indicates that the parameters are fixed when using pre-trained DecL.}
    \label{fig:architecture}
\end{figure*}

MARL employs neural networks for state representation extraction and decision-making, which are tightly coupled.
We train multiple multi-agent tasks simultaneously, where each task utilizes independent front-end network layers alongside shared back-end network layers.
The shared back-end network layers effectively decouple state representation extraction from decision-making, thus facilitating the training of back-end network layers encompassing common decision knowledge.
When addressing new multi-agent tasks, we leverage the back-end network layers containing shared decision knowledge, keeping network parameters fixed and focusing solely on improving the front-end network layers' ability to extract state representations.
Through this approach, we substantially reduce training costs.
The framework of our MTAPN approach is illustrated in Fig.~\ref{fig:architecture}.

\subsection{Multi-Task Pre-training}

We refer to the front-end network layer as the Perception Layer (PerL) and the back-end network layer as the Decision Layer (DecL).
During the pre-training phase, we simultaneously train $N$ multi-agent tasks.
For each $task_{n}$,~$n=\{1,2,\cdots, N\}$, we collect data generated by the interaction between agents and the environment in $task_{n}$ and store it in the replay buffer $Buffer_{n}$.
We then sample from $Buffer_{n}$ for training.
Samples are input into ${\rm PerL}_{n}$ to extract perceptual features, followed by using DecL to predict $action_{n}$.
Each $task_{n}$ has its dedicated ${\rm PerL}_{n}$ while sharing DecL.
Once pre-training is complete, we obtain DecL containing common decision knowledge among different tasks.
We fix the parameters of the pre-trained DecL and transfer it to a new task, subsequently training PerL for that task.
This approach efficiently allows us to learn effective policies for new tasks, leading to a significant reduction in training costs.


\subsection{Action Prepositioning Network}

Value-based MARL algorithms, such as VDN and QMIX, estimate state-action values by taking the state $s$ as the network input and generating estimated values for each action $a_{i}$ under that state. The network's output size remains fixed at the size of the task's action space. Due to variations in action spaces across different multi-agent tasks, sharing the back-end network layer between tasks is infeasible. To address this challenge, we position the action $a_{i}$ at the forefront of the network's input. The network takes state-action pairs as input and yields corresponding value estimates as output. This structural design maintains a fixed network output size of $1$. By utilizing front-end network layers to map inputs from different tasks to a consistent dimension $d$, we enable sharing the back-end network layer across different tasks. Pairing state and all actions as inputs to the network results in value estimates for all state-action pairs. This network architecture is referred to as the Action Prepositioning Network (APN). 

Based on APN, we can share DecL across different tasks during multi-task pre-training. We will illustrate the training process of any task $task_{n}$ using APN at time step $t$. For simplicity, we omit the $t$ subscripts for all variables. $task_{n}$ comprises $M$ agents, each with an action space size of $K$. $o_{m}$ represents the observation of agent $m$, $a_{m}^{'}$ is the action executed by agent $m$ in the previous time step, $id_{m}$ is the identifier of agent $m$, and $a_{m,k}$ represents the $k$-th executable action of agent $m$,~$m=\{1,2,\cdots,M\}$,~$k=\{1,2,\cdots,K\}$. For each agent $m$, perceptual features $f_{m}$ are extracted from the task's input using independent ${\rm PerL}_{n}$: 

\begin{equation}
    f_{m} = {\rm PerL}_{n}(o_{m}, a_{m}^{'}, id_{m}, a_{m,1}, a_{m,2}, \cdots, a_{m,K}),
\end{equation}

\noindent where $f_{m} \in \mathbb{R}^{K \times d}$, with $d$ representing the dimension of perceptual features. Given that each agent can only partially observe the environment at each time step, we employ the GRU recurrent neural network in PerL to encode historical trajectories and address the issue of partial observability. These $f_{m}$ are inputted into the shared DecL to predict value estimates $Q_{m}$: 

\begin{equation}
    Q_{m} = {\rm DecL}(f_{m}),
\end{equation}

\noindent where $Q_{n} \in \mathbb{R}^{K \times 1}$. We select the maximum Q value, denoted as $Q_{*}$, for each individual agent and aggregate them through a mixer to predict the total Q value $Q_{\rm predict}$. In this context, we employ a mixer the same as VDN.
We input the same data into the target APN network to generate the target total Q value $Q_{\rm target}$.
The APN is updated by minimizing the error between $Q_{\rm predict}$ and $Q_{\rm target}$. 

\subsection{Dynamic Adaptive Weighting for Multi-Task Losses}

During multi-task training, not all tasks converge at the same pace.
Typically, simpler tasks tend to converge faster than complex ones.
To enhance the efficiency of multi-task training, it becomes essential to allocate larger loss weights to those tasks that exhibit slower convergence.
We standardize evaluation metrics across different tasks.
$EM_{n}$ represents the normalized evaluation metric value for $task_{n}$ and is used to assess the current convergence status of $task_{n}$.
$EM_{n}^{-}$ represents the inverse evaluation metric of $EM_{n}$, such as the negative or reciprocal value of $EM_{n}$.
A smaller $EM_{n}$, or equivalently, a larger $EM_{n}^{-}$, indicates poorer convergence for the corresponding task.
We introduce a method for dynamically and adaptively assigning weights to multi-task losses, wherein the weights are adjusted as $EM_{n}^{-}$ change during the training process.
Here, $loss_{n}$ corresponds to the loss of $task_{n}$, $\omega_{n}$ represents the weight of $loss_{n}$, and the total loss $total\_loss$ for all tasks is calculated as follows: 

\begin{align}
    \omega_{n} = EM_{n}^{-}/\sum_{i=1}^{N}EM_{i}^{-},\\
    total\_loss = \sum_{i=1}^{N}(\omega_{i} \times loss_{i}).
\end{align}

\section{Experiments}
\label{sec:exps}

We conducted experiments in the StarCraft Multi-agent Challenge (SMAC)~\cite{samvelyan2019starcraft} and Google Research Football (GRF)~\cite{kurach2020google} environments.
Unless otherwise specified, both the pre-training algorithm and downstream task algorithm utilized in the experiments are VDN~\cite{sunehag2017value}.
Multiple tasks are employed to pre-train the VDN network based on APN, extracting the DecL containing common knowledge.
Utilizing the common knowledge in DecL, we exclusively train the PerL on new tasks.
In SMAC and GRF tasks, we employ win rate as $EM$ and failure rate as $EM^{-}$.
Each set of experimental results is obtained using $5$ random seeds.
The solid line shows the mean win rate, and the shaded area represents the minimum to maximum win rate for 5 random seeds. 
Actions are chosen for exploration based on the estimated $Q$ values using an $\epsilon$-greedy policy.
The algorithms we use are based on the Pymarl2 algorithm library~\cite{hu2021rethinking}.
Detailed hyperparameter settings and code are provided in the supplementary material.

\subsection{StarCraft Multi-Agent Challenge (SMAC)}

We increase the number of tasks used for pre-training to explore whether incorporating more tasks can lead to improved acquisition of common knowledge.
All our experiments were conducted specifically on the hard and super-hard difficulty levels of the SMAC task.
The tasks involved in the experiments are detailed in table~\ref{smac maps}.

\begin{table}
\centering
\caption{The maps in the SMAC tasks used in the experiment and their corresponding difficulty.}
\label{smac maps}
\begin{tabular}{ccc} 
\toprule
Task   & Map            & Difficulty  \\ 
\hline
$task_{1}$  & 5m\_vs\_6m     & Hard        \\
$task_{2}$  & 8m\_vs\_9m     & Hard        \\
$task_{3}$  & 3s\_vs\_5z     & Hard        \\
$task_{4}$  & bane\_vs\_bane & Hard        \\
$task_{5}$  & 2c\_vs\_64zg   & Hard        \\
$task_{6}$  & corridor       & Super Hard  \\
$task_{7}$  & MMM2           & Super Hard  \\
$task_{8}$  & 3s5z\_vs\_3s6z & Super Hard  \\
$task_{9}$  & 27m\_vs\_30m   & Super Hard  \\
$task_{10}$ & 6h\_vs\_8z     & Super Hard  \\
\bottomrule
\end{tabular}
\end{table}

\begin{figure*}[thbp!]
    \centering
    \begin{tabular}{@{\extracolsep{\fill}}c@{}c@{}c@{}@{\extracolsep{\fill}}}
            \includegraphics[width=0.33\linewidth]{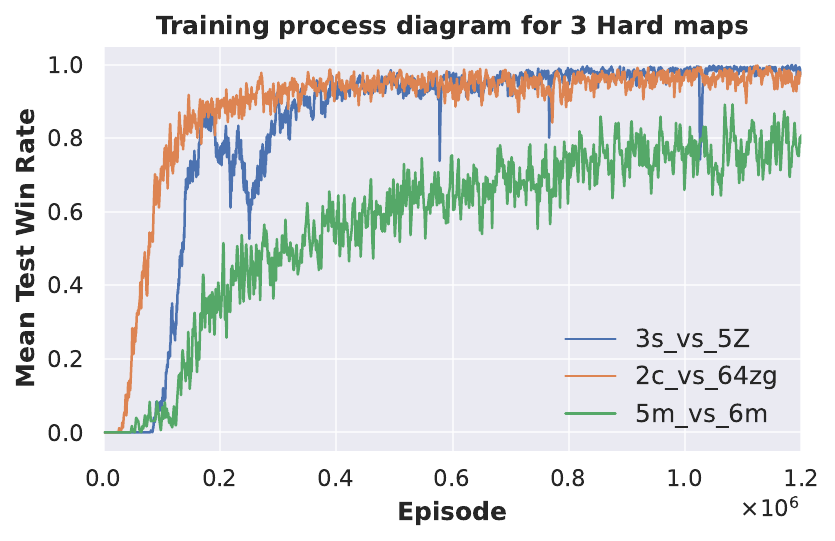} & 
            \includegraphics[width=0.33\linewidth]{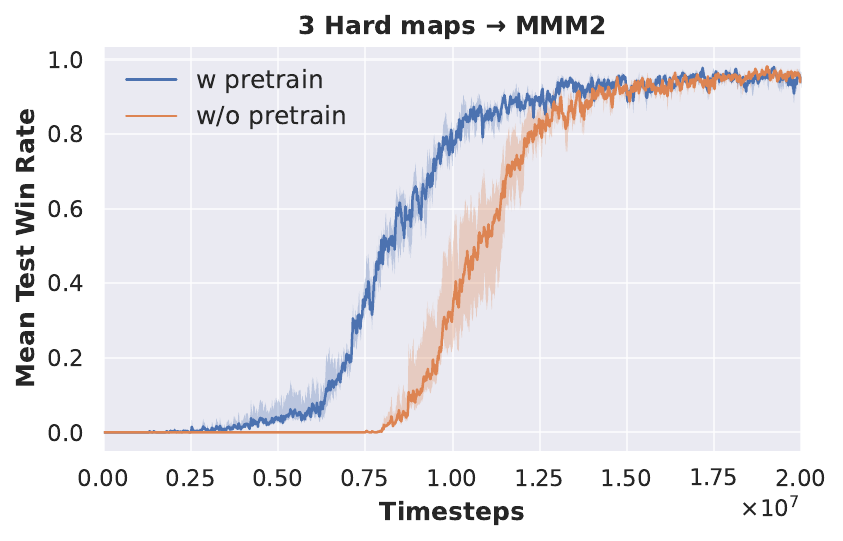} &
            \includegraphics[width=0.33\linewidth]{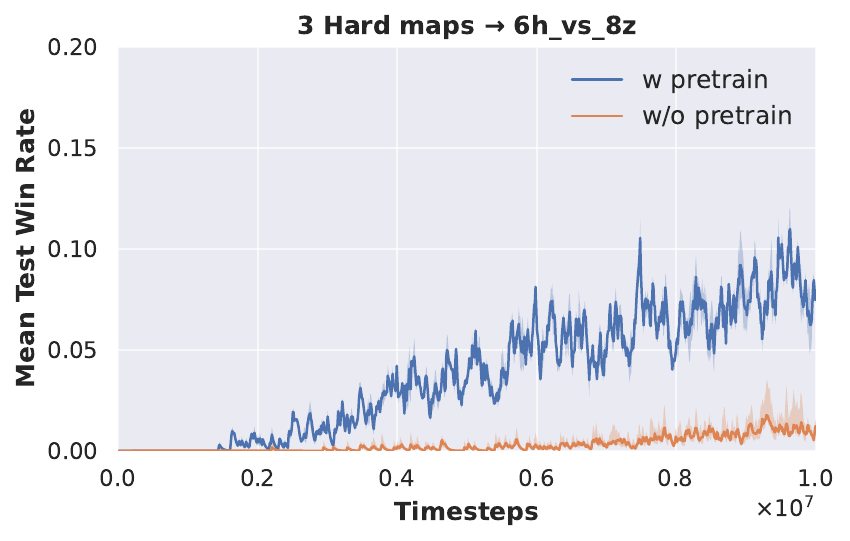} \\
            (a) pre-training of $Group_{1}$ tasks & (b) from $Group_{1}$ tasks to $task_{7}$ &
            (c) from $Group_{1}$ tasks to $task_{10}$\\
    \end{tabular}
    \begin{tabular}{@{\extracolsep{\fill}}c@{}c@{}c@{}@{\extracolsep{\fill}}}
            \includegraphics[width=0.33\linewidth]{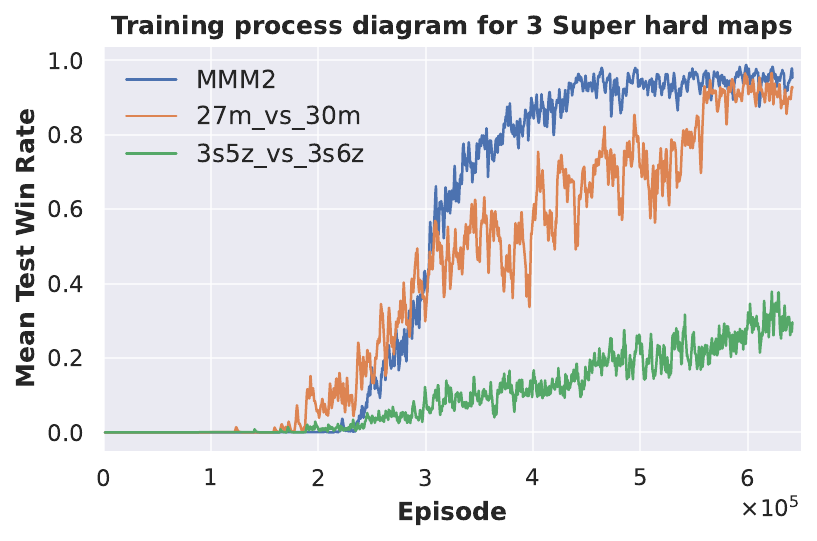} & 
            \includegraphics[width=0.33\linewidth]{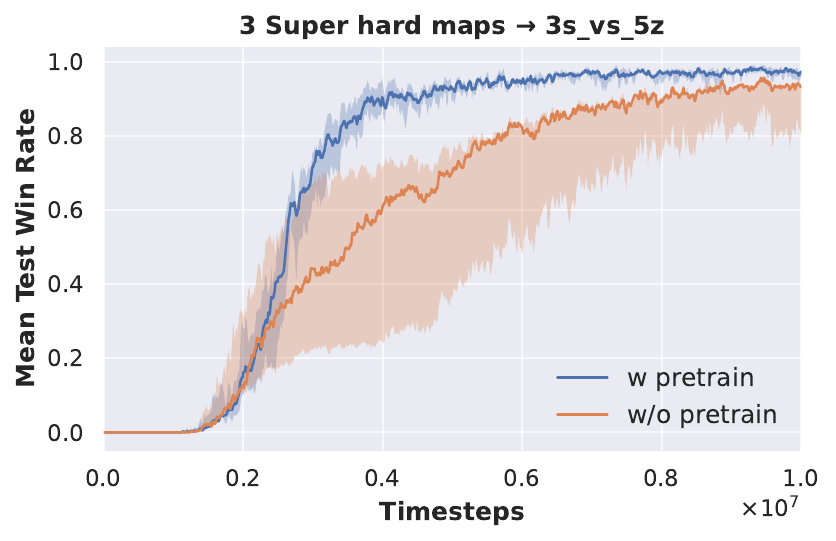} &
            \includegraphics[width=0.33\linewidth]{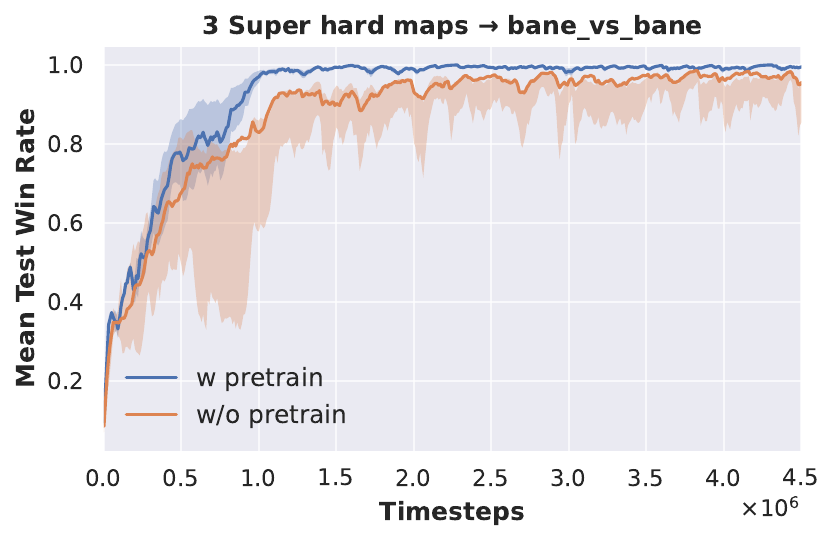} \\
            (d) pre-training of $Group_{2}$ tasks& (e) from $Group_{2}$ tasks to $task_{3}$ &
            (f) from $Group_{2}$ tasks to $task_{4}$
    \end{tabular}
    \caption{Win rates of DecL pre-training using three tasks and transferred to the new SMAC task.}
    \label{fig:3 to 1}
\end{figure*}

\noindent \textbf{Three-task pre-training.} We employ two sets of tasks for pre-training DecL: $Group_{1}: (task_{1}, task_{3}, task_{5})$ and $Group_{2}: (task_{7},\\ task_{8}, task_{9})$.
All tasks within $Group_{1}$ are of hard difficulty, while those in $Group_{2}$ are of super hard difficulty.
The DecL pre-trained with $Group_{1}$ is transferred and separately fine-tuned for PerL training on $task_{7}$ and $task_{10}$.
Similarly, the DecL pre-trained with $Group_{2}$ is transferred and separately fine-tuned for PerL training on $task_{3}$ and $task_{4}$. Experimental results are depicted in Fig.~\ref{fig:3 to 1}.

From Fig.~\ref{fig:3 to 1} (a) and (b), it is evident that the APN network, during multi-task training, effectively learns strategies for each task. The DecL obtained from multi-task pre-training encompasses decision knowledge utilized in all tasks, namely, generic decision knowledge. Fig.~\ref{fig:3 to 1} (c) shows that using the DecL pre-trained on three hard tasks results in faster convergence for $task_{7}$ compared to training from scratch. Fig.~\ref{fig:3 to 1} (d) demonstrates that, with the DecL pre-trained on three hard tasks, $task_{10}$ achieves superior convergence compared to starting from scratch. Fig.~\ref{fig:3 to 1} (e) shows that utilizing the DecL pre-trained on three super hard tasks leads to faster convergence and better final convergence results for $task_{3}$ compared to starting from scratch. However, Fig.~\ref{fig:3 to 1} (f) reveals that, despite being pre-trained on three super hard tasks, $task_{4}$ exhibits convergence results and rates consistent with training from scratch, indicating no improvement. Clearly, in comparison to DecL pre-trained on a single task, DecL pre-trained on three tasks shows better performance when transferring to new tasks. This highlights the advantage of having more pre-training tasks in obtaining DecL with more comprehensive generic decision knowledge. The lack of improvement in the convergence performance of $task_{4}$ also suggests that the decision knowledge contained within DecL can only cover some tasks comprehensively, prompting further consideration of increasing the number of pre-training tasks.

\begin{figure*}[thbp!]
    \centering
    \begin{tabular}{@{\extracolsep{\fill}}c@{}c@{}c@{}@{\extracolsep{\fill}}}
            \includegraphics[width=0.33\linewidth]{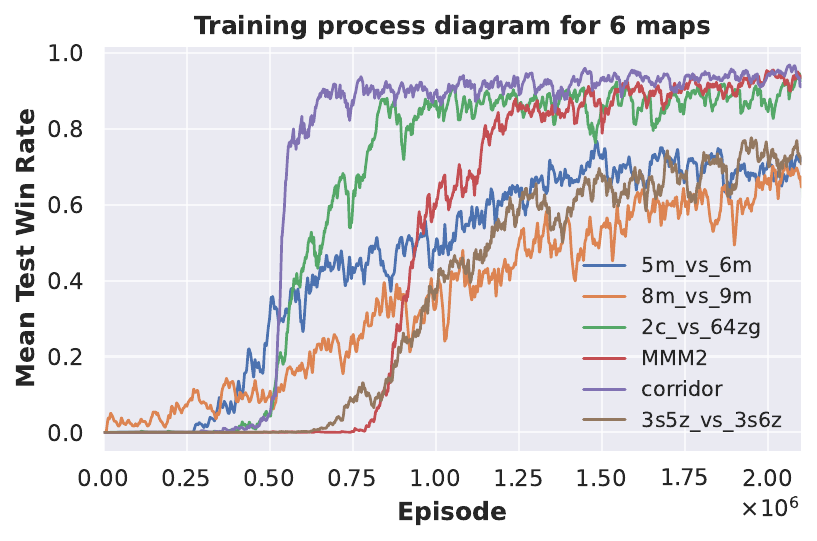} & 
            \includegraphics[width=0.33\linewidth]{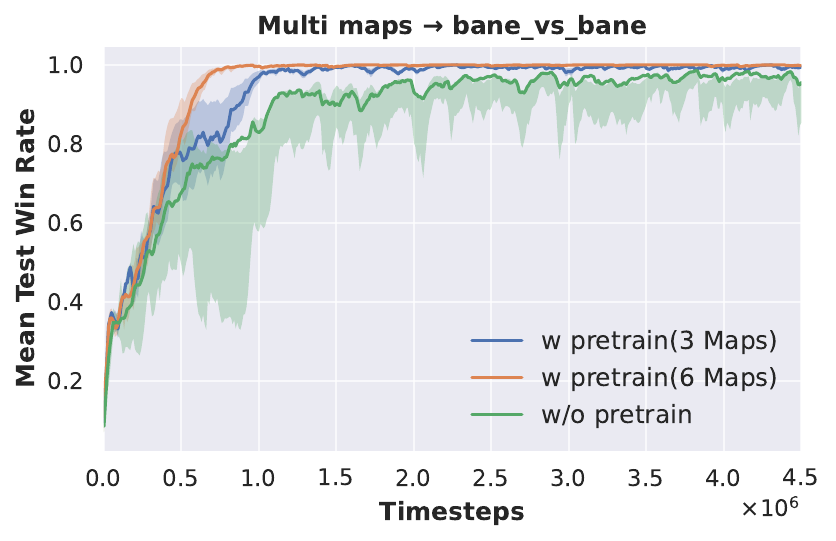} &
            \includegraphics[width=0.33\linewidth]{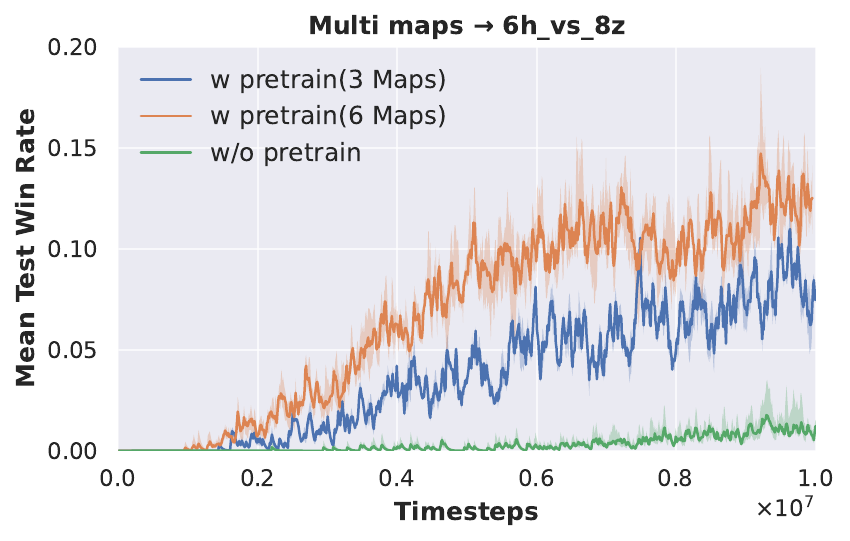} \\
            (a) pre-training of $Group_{3}$ tasks & (b) from $Group_{3}$ tasks to $task_{4}$ &
            (c) from $Group_{3}$ tasks to $task_{10}$\\
    \end{tabular}
    \caption{Win rates of $Group_{3}$ tasks pre-training, and using $Group_{3}$ pre-trained DecL to train $task_{4}$ and $task_{10}$.}
    \label{fig:smac 6 to 1}
\end{figure*}

\noindent \textbf{Six-task pre-training.} We select six tasks for pre-training DecL:\\$Group_{3}:(task_{1}, task_{2}, task_{5}, task_{6}, task_{7}, task_{8})$. We fix the pre-trained DecL parameters and transfer them separately to $task_{4}$ and $task_{10}$. The experimental results are depicted in Fig.~\ref{fig:smac 6 to 1}. 

Fig.~\ref{fig:smac 6 to 1} (a) demonstrates that our proposed method continues to acquire effective policies across all tasks when the number of pre-training tasks increases to six. In Fig.~\ref{fig:smac 6 to 1} (b), we observe that training $task_{4}$ with DecL pre-trained based on $Group_{3}$ tasks results in faster convergence and superior final performance compared to training from scratch. Furthermore, employing DecL pre-trained on $Group_{3}$ tasks for $task_{4}$ converges more rapidly than using DecL pre-trained on $Group_{2}$ tasks. The same result can also be found in Fig.~\ref{fig:smac 6 to 1} (c). Training $task_{10}$ with DecL pre-trained on $Group_{3}$ leads to faster convergence and improved outcomes compared to using $Group_{1}$ pre-trained DecL. In summary, through multi-task pre-training, we can extract DecL containing common decision knowledge, which becomes more generalized with increased pre-training tasks.

\begin{figure}[thbp!]
    \centering
    \begin{tabular}{@{\extracolsep{\fill}}c@{}c@{}@{\extracolsep{\fill}}}
            \includegraphics[width=0.33\linewidth]{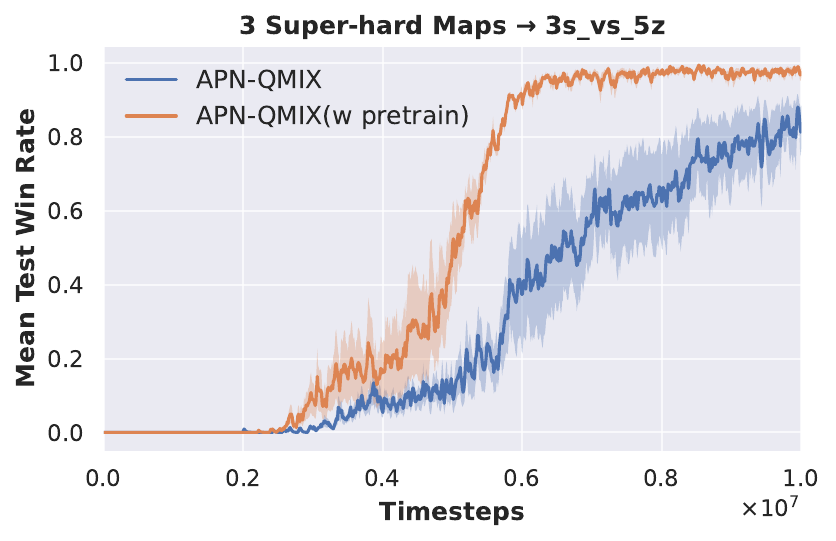} &
            \includegraphics[width=0.33\linewidth]{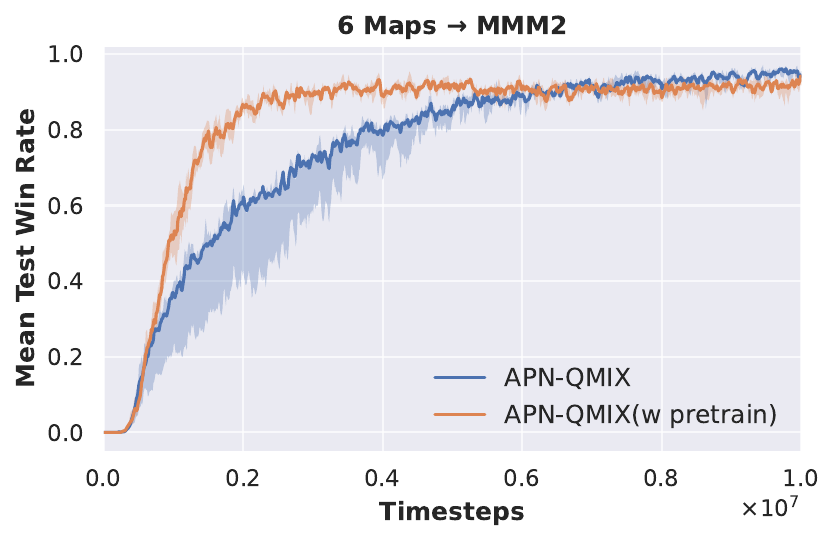}\\
            (a) from $Group_{2}$ tasks to $task_{3}$ & (b) from $Group_{1}$ tasks to $task_{7}$
    \end{tabular}
    \caption{Win rates of Using QMIX to train new tasks with DecL pre-trained on multi-tasks based on APN-VDN.}
    \label{fig:vdn to qmix}
\end{figure}

\noindent \textbf{VDN to QMIX.} To further validate the inclusion of common decision knowledge in the DecL pre-trained with the VDN algorithm, we migrate the pre-trained DecL from three hard difficulty and three super hard difficulty SAMC tasks to $task_{3}$ and $task_{7}$, respectively. We employ the QMIX algorithm to train PerL on these new tasks. Experimental results are presented in Fig.~\ref{fig:vdn to qmix}.

Fig.~\ref{fig:vdn to qmix} demonstrates that when the DecL pre-trained on three tasks using the VDN algorithm is transferred to new tasks and trained with the QMIX algorithm, the final convergence results are superior to those of the QMIX algorithm trained from scratch.
Consequently, the DecL obtained from multi-task pre-training indeed encompasses common decision knowledge.
This common decision knowledge is applicable to various new tasks as well as new multi-agent reinforcement learning algorithms.

\subsection{Ablation Study}

\begin{figure*}[thbp!]
    \centering
    \begin{tabular}{@{\extracolsep{\fill}}c@{}c@{}c@{}@{\extracolsep{\fill}}}
            \includegraphics[width=0.33\linewidth]{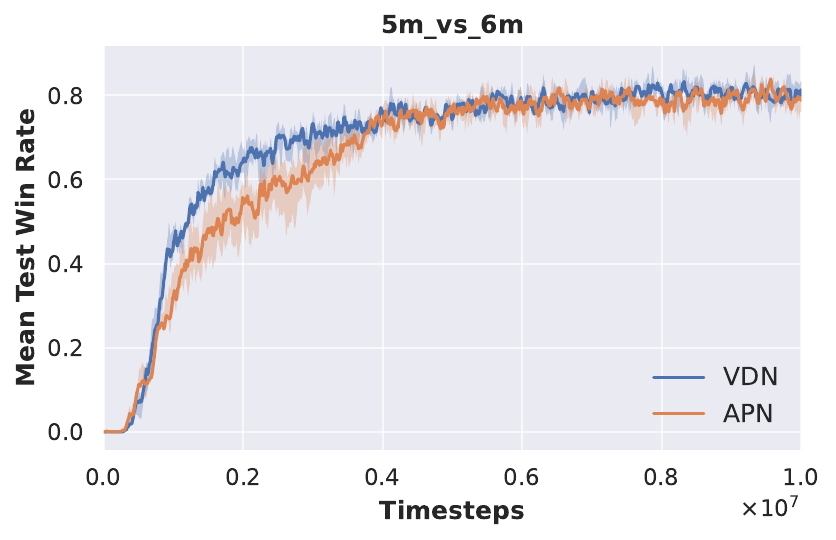} & 
            \includegraphics[width=0.33\linewidth]{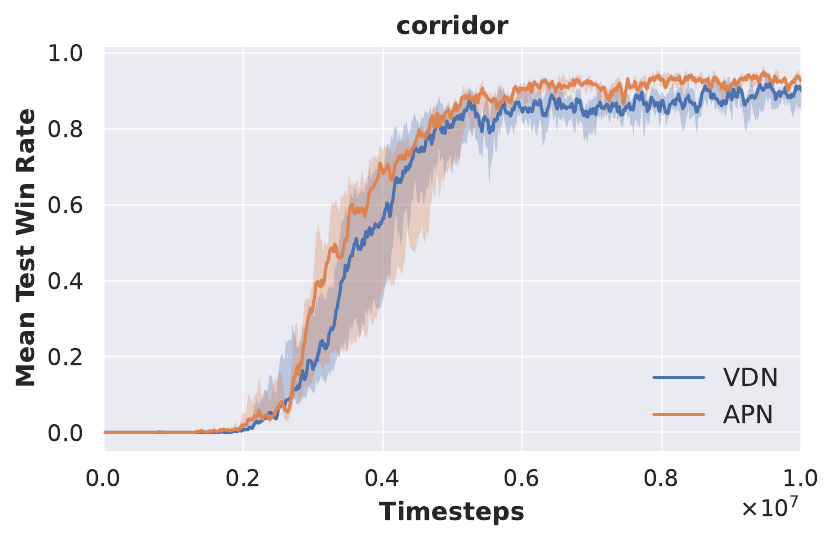} &
            \includegraphics[width=0.33\linewidth]{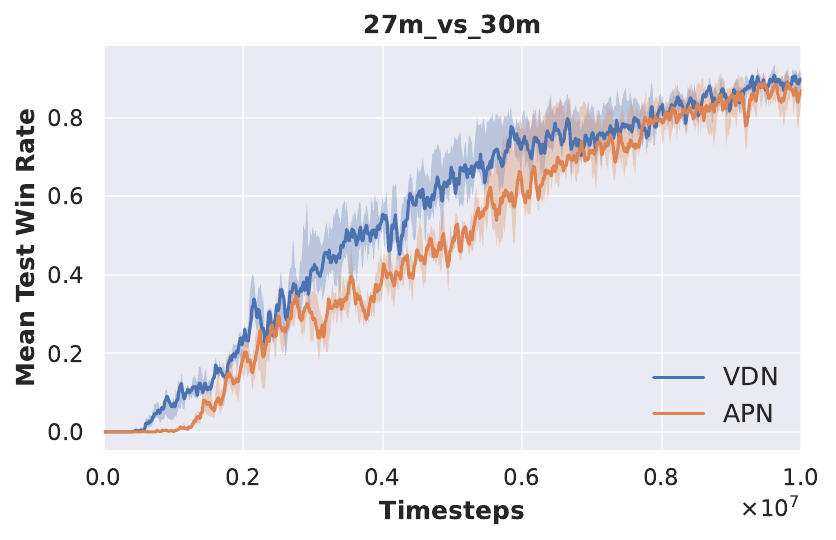} \\
            (a) $task_{1}$ & (b) $task_{6}$ & (c) $task_{9}$\\
    \end{tabular}
    \caption{Win rates of training SMAC tasks using original VDN and APN-VDN.}
    \label{fig:vdn vs apn}
\end{figure*}

\textbf{APN-VDN vs VDN.} To enable multi-task training, we initially place the action at the front of the network input and maintain this structure when transitioning to new tasks. 
To illustrate the impact of adopting this structure on experimental performance, we train original VDN and APN-VDN on all SMAC tasks, with the results presented in Fig.~\ref{fig:vdn vs apn} and table~\ref{vdn vs apn}.

\begin{table}
\centering
\caption{Convergence results of original VDN and APN-VDN on all SMAC tasks. Bold indicates optimal results. Following $``\pm"$ is the variance of the experimental results of multiple random seeds.}
\label{vdn vs apn}
\begin{tabular}{cccc} 
\toprule
Map            & Difficulty & original VDN        & APN-VDN         \\ 
\hline
5m\_vs\_6m     & Hard       & \textbf{80.38$\pm$0.06} & 79.31$\pm$0.06  \\
8m\_vs\_9m     & Hard       & \textbf{89.37$\pm$0.08} & 86.55$\pm$0.05  \\
3s\_vs\_5z     & Hard       & 94.93$\pm$0.11 & \textbf{98.05$\pm$0.01}  \\
bane\_vs\_bane & Hard       & 97.16$\pm$0.02 & \textbf{99.97$\pm$0.00}  \\
2c\_vs\_64zg   & Hard       & \textbf{97.24$\pm$0.01} & 96.11$\pm$0.05  \\
corridor       & Super Hard & 90.74$\pm$0.09  & \textbf{92.45$\pm$0.03} \\
MMM2           & Super Hard & 96.27$\pm$0.01 & \textbf{96.99$\pm$0.01}  \\
3s5z\_vs\_3s6z & Super Hard & \textbf{73.65$\pm$0.17} & 62.11$\pm$0.17  \\
27m\_vs\_30m   & Super Hard & 95.99$\pm$0.01 & \textbf{97.71$\pm$0.00}  \\
6h\_vs\_8z     & Super Hard & 1.08$\pm$0.01  & \textbf{1.83$\pm$0.01}   \\
\bottomrule
\end{tabular}
\end{table}

As shown in Fig.~\ref{fig:vdn vs apn}, adopting the APN structure has a slight impact on the convergence speed of the original VDN but does not affect the ultimate convergence results.
Table~\ref{vdn vs apn} demonstrates that APN-VDN and original VDN achieve nearly identical convergence results across all SMAC tasks.
Therefore, employing the APN structure does not influence the performance of multi-agent algorithms.
Consequently, we use APN-VDN as the baseline algorithm for comparison in our experiments, with variations in algorithm performance stemming from the utilization of pre-trained DecL.

\begin{figure}[thbp!]
    \centering
    \begin{tabular}{@{\extracolsep{\fill}}c@{}c@{}@{\extracolsep{\fill}}}
            \includegraphics[width=0.33\linewidth]{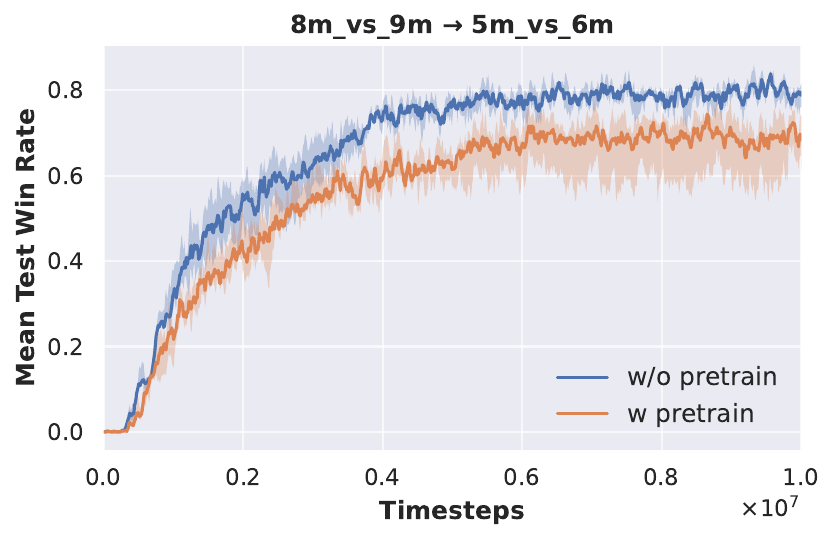} & 
            \includegraphics[width=0.33\linewidth]{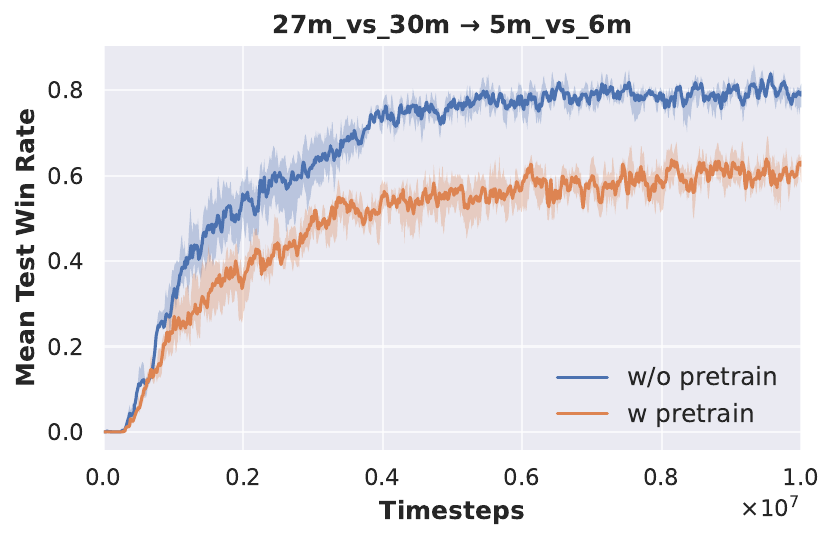}\\
            (a) from $task_{2}$ to $task_{1}$ & (b) from $task_{9}$ to $task_{1}$
    \end{tabular}
    \caption{Win rates of training new SMAC tasks using single-task pre-trained DecL.}
    \label{fig:1 to 1}
\end{figure}

\noindent  \textbf{Single-task Pre-training.} We conduct experiments on $task_{1}$, $task_{2}$, and $task_{9}$.
We perform pre-training on $task_{2}$ and $task_{9}$ to obtain DecL with task-specific decision knowledge.
Subsequently, we freeze the parameters of DecL and transfer them to $task_{1}$ for retraining PerL.
Experimental results are illustrated in Fig.~\ref{fig:1 to 1}.

Fig.~\ref{fig:1 to 1} reveals that when using DecL pre-trained on $task_{2}$ and $task_{9}$, the final convergence outcomes for $task_{1}$ are consistently worse compared to training from scratch.
This occurs because pre-training DecL on a single task leads to overfitting on that specific task, preventing the acquisition of general decision knowledge.
Consequently, when we transfer a DecL pre-trained on a single task to a new task, it fails to align with the new task, resulting in a deterioration of the final convergence performance for the new task.
Therefore, we need to use more pre-training tasks to extract DecL containing general decision-making knowledge.

\begin{figure}[thbp!]
    \centering
    \includegraphics[width=0.5\linewidth]{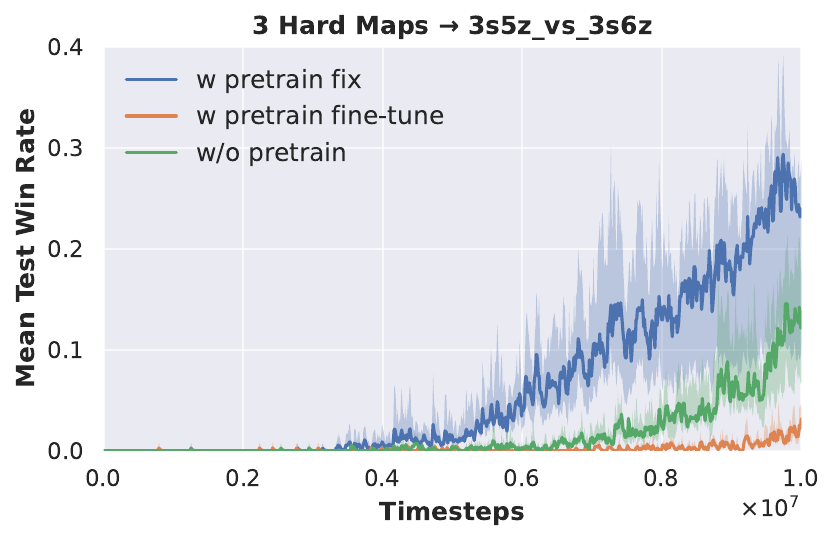}
    \caption{Win rates of training $task_{8}$ when using a fixed pre-trained DecL, fine-tuning the pre-trained DecL, and not using the pre-trained DecL. Pre-trained DecL is obtained by pre-training on the $Group_{1}$ tasks.}
    \label{fig:fix vs finetune}
\end{figure}

\noindent \textbf{Fix vs Fine-tune.} When employing pre-trained models, we typically face the choice between weight fixation and fine-tuning. Our experiments explore both scenarios when transferring pre-trained DecL weights on $Group_{1}$ tasks to $task_{8}$.
The experimental results, as depicted in Fig.~\ref{fig:fix vs finetune}, demonstrate that fixing the pre-trained DecL weights leads to efficient convergence to superior results.
During the backpropagation process, the loss diminishes as it propagates from the network's output to the input.
DecL is closer to the network's output, resulting in more significant parameter updates compared to PerL.
In the early stages of training, PerL has not yet developed robust perceptual capabilities.
Consequently, the transmission of substantial losses can disrupt the pre-trained DecL, leading to decreased convergence speed and inferior results. Thus, we consistently opt for the fixed-weight approach when employing pre-trained DecL.

\begin{figure*}[thbp!]
    \centering
    \begin{tabular}{@{\extracolsep{\fill}}c@{\hskip 0.75em}c@{\hskip 0.75em}c@{}@{\extracolsep{\fill}}}
            \includegraphics[width=0.32\linewidth]{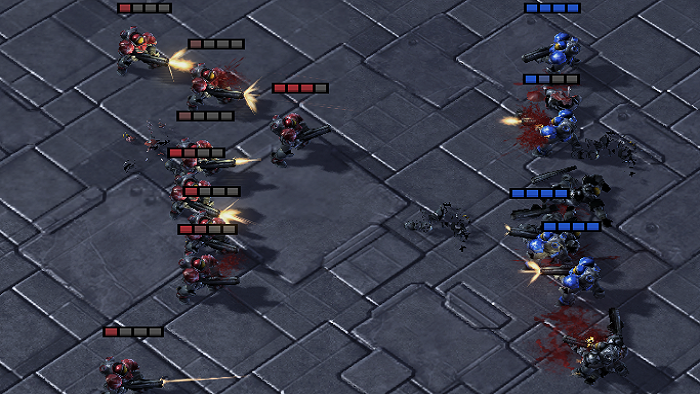} & 
            \includegraphics[width=0.32\linewidth]{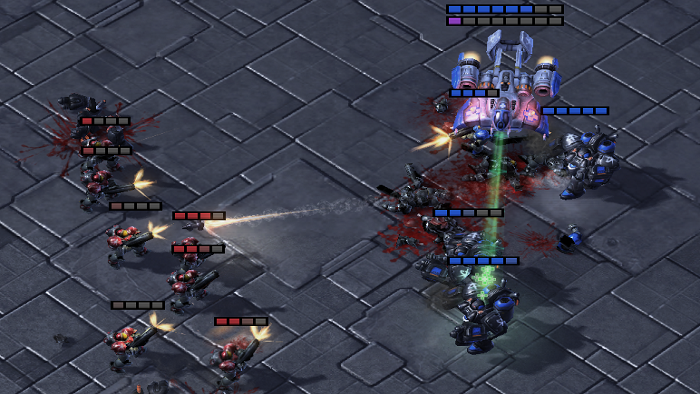} &
            \includegraphics[width=0.32\linewidth]{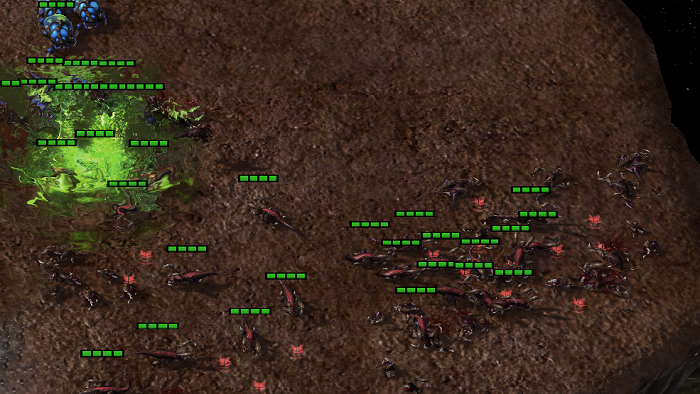} \\
            (a) $task_{2}$ w/o pre-trained DecL at $6$M steps & (b) $task_{7}$ w/o pre-trained DecL at $10$M steps & (c) $task_{4}$ w/o pre-trained DecL at $0.6$M steps\\
             \includegraphics[width=0.32\linewidth]{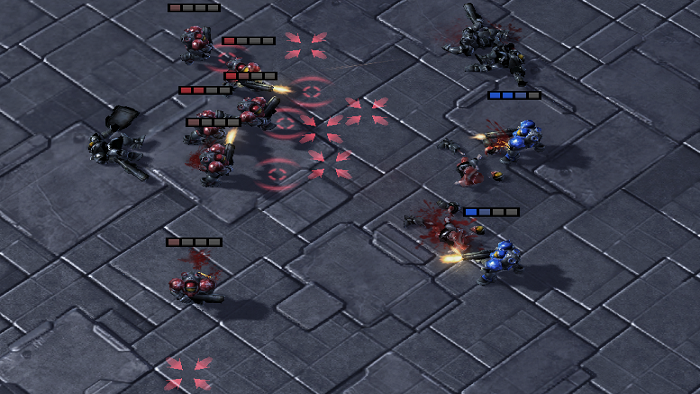} & 
            \includegraphics[width=0.32\linewidth]{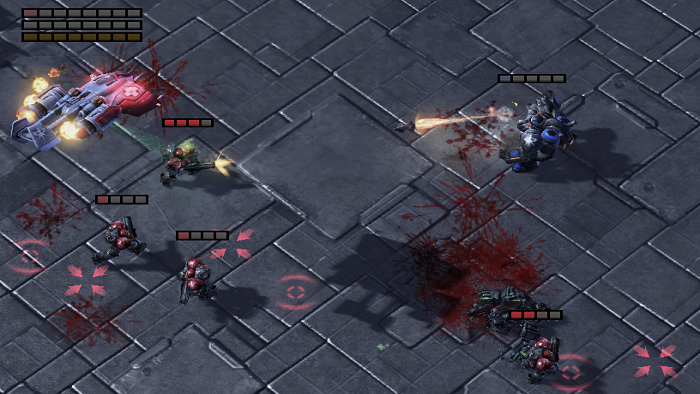} &
            \includegraphics[width=0.32\linewidth]{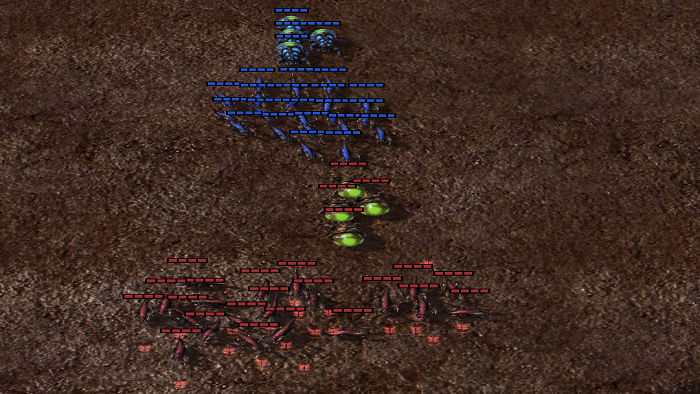} \\
            (d) $task_{2}$ with pre-trained DecL at $6$M steps & (e) $task_{7}$ with pre-trained DecL at $10$M steps & (f) $task_{4}$ with pre-trained DecL at $0.6$M steps\\
    \end{tabular}
    \caption{Visualization of model performance for different tasks with and without pre-trained decision modules at the same training steps.}
    \label{fig:visual}
\end{figure*}

\subsection{Visualization}

To better illustrate how our proposed multi-task multi-agent reinforcement learning algorithm learns to share common decision knowledge between different tasks, we visualize the acquired policies in the SMAC tasks.
The videos of the battle are provided in the supplementary material.

In Fig.~\ref{fig:visual}(a) and (d), we visualize the performance on $task_{2}$ after training for $6$ million steps with and without the pre-trained DecL using $Group_{2}$ tasks. In Fig.~\ref{fig:visual}(a), the attack positions of all Marines remain nearly unchanged, displaying a lack of flexibility, even when their health is low. However, in Fig.~\ref{fig:visual}(d), the Marines exhibit a more flexible policy by moving to disperse enemy fire and concentrating their attacks on a portion of the enemy forces.

Fig.~\ref{fig:visual}(b) and (e) visualize the performance on $task_{7}$ after training for $10$ million steps without pre-trained DecL and with DecL pre-trained on $Group_{1}$ tasks, respectively. In Fig.~\ref{fig:visual}(b), the Medivac does not employ an evasive strategy, making it vulnerable to enemy attacks, resulting in losses to our Medivac, while the health of other agents cannot recover promptly. Additionally, the positioning of Marines and Marauders lacks strategic depth, failing to execute timely retreats when their health is low. However, in Fig.~\ref{fig:visual}(e), the agents have acquired advanced policies: Marines with low health retreat promptly, while healthier Marines draw enemy Marauders' fire, allowing other Marines to flank the enemy Marauders for surprise attacks stealthily.

In Fig.~\ref{fig:visual}(c) and (f), we visualize the performance on $task_{4}$ after training for $0.6$ million steps, comparing scenarios with and without pre-trained DecL from $Group_{3}$ tasks. In Fig.~\ref{fig:visual}(c), all Banelings charge towards the opponent to detonate them, while some Zerglings also advance but are damaged by the Banelings' explosions. However, in Fig.~\ref{fig:visual}(f), all Banelings immediately rush towards the opposing camp, while Zerglings quickly retreat to avoid unintentional harm from the Banelings, resulting in minimal casualties.

In summary, by utilizing a pre-trained DecL incorporating generic decision knowledge to train for new tasks, we typically acquire more effective policies within the same training steps, thus achieving superior results more efficiently.

    
    

\subsection{Google Research Football (GRF)}

\begin{figure*}[thbp!]
    \centering
    \begin{tabular}{@{\extracolsep{\fill}}c@{}c@{}c@{}@{\extracolsep{\fill}}}
            \includegraphics[width=0.33\linewidth]{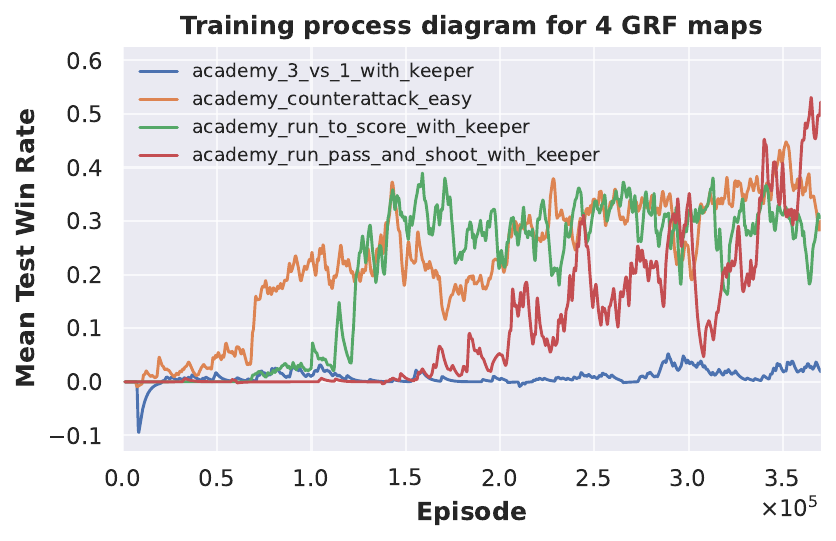} & 
            \includegraphics[width=0.33\linewidth]{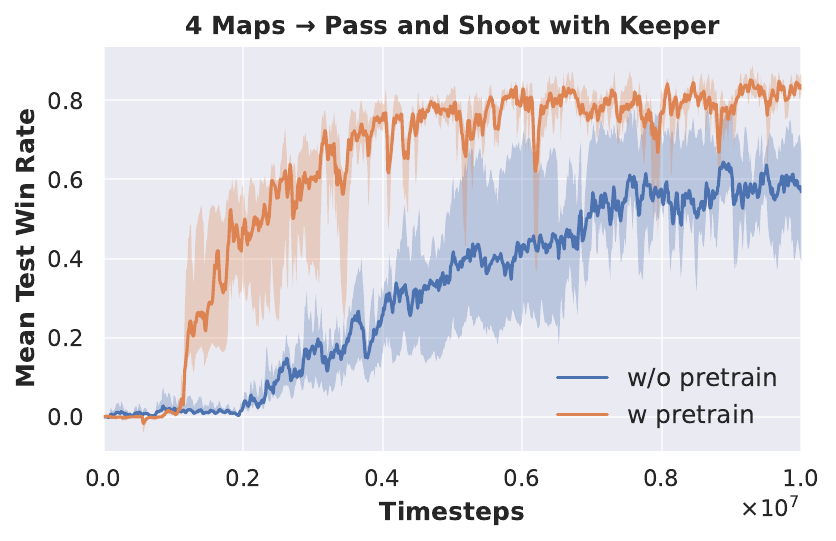} &
            \includegraphics[width=0.33\linewidth]{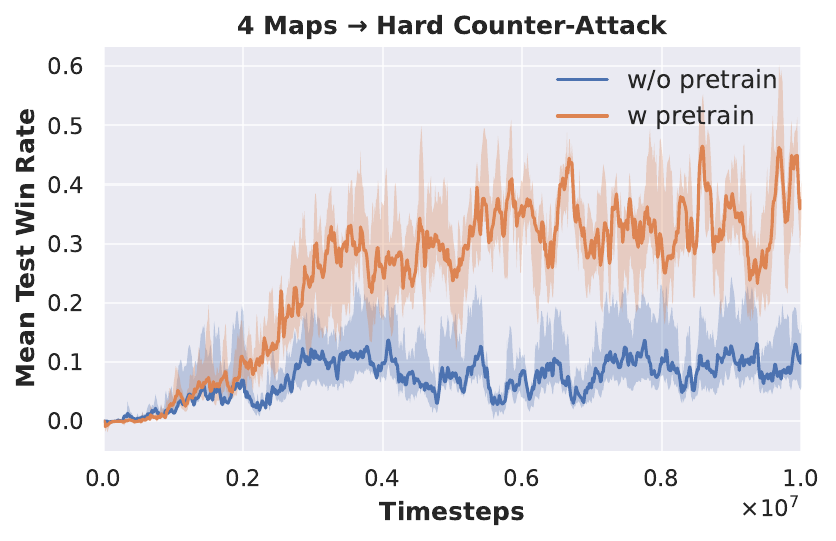} \\
            (a) pre-training of 4 GRF tasks & (b) from four tasks to $task_{12}$ &
            (c) from four tasks to $task_{16}$\\
    \end{tabular}
    \caption{Win rates of ``pre-training DecL using four GRF tasks" and ``transferring DecL pre-trained on GRF tasks to the new GRF task".}
    \label{fig:4 to 1}
\end{figure*}

We conduct experiments on six tasks from GRF, as shown in table~\ref{grf maps}. We choose four tasks, including both easy and hard difficulties $(task_{11}, task_{13}, task_{14}, task_{15})$, for pre-training the DecL. Subsequently, we transfer the pre-trained DecL to $(task_{12}$ of easy difficulty and $(task_{16}$ of super hard difficulty. The experimental outcomes are depicted in Fig.~\ref{fig:4 to 1}.

\begin{table}
\centering
\caption{The maps in the GRF task used in the experiment and their corresponding difficulty.}
\label{grf maps}
\begin{tabular}{ccc} 
\toprule
Task       & Map                             & Difficulty    \\ 
\hline
$task_{11}$ & Run to Score with Keeper        & Easy          \\
$task_{12}$ & Pass and Shoot with Keeper      & Easy          \\
$task_{13}$ & Run, Pass and Shoot with Keeper & Easy          \\
$task_{14}$ & 3 versus 1 with Keeper          & Hard          \\
$task_{15}$ & Easy Counter-Attack             & Hard          \\
$task_{16}$ & Hard Counter-Attack             & Super Hard    \\
\bottomrule
\end{tabular}
\end{table}

Fig.~\ref{fig:4 to 1} (a) illustrates that the APN network, during multi-task pre-training, effectively learns policies across various GRF tasks, each involving a different number of agents. Fig.~\ref{fig:4 to 1} (b) demonstrates that with the pre-trained DecL, the convergence speed and final convergence results for  $task_{12}$ of easy difficulty significantly improve compared to training from scratch. Fig.~\ref{fig:4 to 1} (c) reveals that, with the pre-trained DecL, the final convergence results for $task_{16}$ of super hard difficulty are superior to those obtained from training from scratch. The utilization of pre-trained DecL renders training on new GRF tasks more efficient. Notably, $task_{16}$ fails to acquire effective policy when trained from scratch. In summary, through multi-task pre-training, we can obtain a DecL containing common decision knowledge, which enables us to more efficiently acquire policies for new tasks, substantially reducing training costs.

\section{Conclusion}
\label{sec:conclus}

In this paper, we propose a multi-task multi-agent reinforcement learning algorithm that aims to obtain shared decision modules through multi-task pre-training.
To ensure consistency in the format of shared knowledge across tasks and action spaces, we introduce the action repositioning network (APN).
By relocating actions from the output of the decision-making module to the input of the perception module, we maintain a consistent output dimension for the decision-making module across various agents and tasks.
Additionally, we address the challenge of disparate training speeds among different tasks during multi-task training by dynamically adjusting task weights, resulting in accelerated pre-training.

Through extensive experiments, we demonstrate that the pre-trained shared decision-making module contains general multi-agent decision-making knowledge.
Moreover, as we increase the number of tasks used for pre-training, we observe that the shared decision layer contains more general decision knowledge, further improving the training efficiency for new tasks.
By leveraging a decision module that incorporates this general decision knowledge, we significantly reduce the training cost for new tasks.

In future work, we plan to expand the number of pre-training tasks and train a larger decision-making module that encompasses a broader spectrum of general decision-making knowledge.
Additionally, we aim to explore the transferability of decision-making knowledge across different domains, such as applying knowledge from StarCraft to tasks in Dota or Honor of Kings.
Given the existence of common decision-making knowledge between these tasks, we strongly believe this idea holds considerable promise for future research.


\bibliographystyle{unsrt}  
\bibliography{references}

\end{document}